\documentclass[epj]{webofc}
\usepackage[varg]{txfonts}   
\newcommand{\eqb}{\begin{equation}}
\newcommand{\eqe}{\end{equation}}
\newcommand{\dmb}{\begin{displaymath}}
\newcommand{\dme}{\end{displaymath}}
\newcommand{\pd}{\partial}

\newcommand{\eab}{\begin{eqnarray}}
\newcommand{\eae}{\end{eqnarray}}

\newcommand{\e}{\mbox{e}}
\newcommand{\be}{\begin{equation}}
\newcommand{\ee}{\end{equation}}
\newcommand{\sgn}{\text{sgn}\,}

\wocname{EPJ Web of Conferences}
\woctitle{ICNFP 2017}
\begin{document}
\selectlanguage{english}
\title{SU(2) Yang-Mills thermodynamics:\\ a priori estimate and radiative corrections}
%
%

\author{Ralf Hofmann\inst{1}\fnsep\thanks{\email{r.hofmann@thphys.uni-heidelberg.de}}
}

\institute{Institut f\"ur Theoretische Physik, Universit\"at Heidelberg, Philosophenweg 16, 69120 Heidelberg}

\abstract{%
We review and explain essential characteristics of the a priori estimate of the thermal ground state and 
its excitations in the deconfining 
phase of SU(2) Quantum Yang-Mills thermodynamics. This includes the spatially 
central and peripheral structure of Harrington-Shepard (anti)calorons, a sketch on how a spatial coarse-graining 
over (anti)caloron centers yields an inert scalar field, which is responsible for an adjoint Higgs mechanism, 
the identification of (anti)caloron action with $\hbar$, a discussion of how, owing to (anti)caloron structure, the 
thermal ground state can be excited (wave-like and particle-like massless modes, 
massive thermal quasiparticle fluctuations), the principle role of and accounting for radiative corrections, the 
exclusion of energy-sign combinations due to constraints on momenta transfers in four-vertices in 
a completely fixed, physical gauge, dihedral diagrams and their resummation up to infinite loop order 
in the massive sector, and the resummation of the one-loop polarisation tensor of the massless modes. 
We also outline applications of deconfining SU(2) Yang-Mills thermodynamics 
to the Cosmic Microwave Background (CMB) which affect the cosmological model at high redshifts, the redshift 
for re-ionization of the Universe, the CMB angular power spectra at low $l$, and the late-time emergence of 
intergalactic magnetic fields.      
}
\maketitle
\section{Introduction}
\label{intro}

Non-Abelian gauge theory in four spacetime dimensions exhibits beautiful, rich, 
and deep structures with clear and convincing links to experimentally accessible, fundamental phenomena such as asymptotic freedom \cite{GrossWilczek,Politzer,KryplovichI,KryplovichII,tHooft1}, chiral symmetry breaking \cite{tHooft2,BanksCasher}, 
confinement \cite{tHooft3,Mandelstam}, the axial anomaly \cite{AdlerBardeen,Jackiw,Fujikawa}, 
and fundamental, theoretically prescribed \cite{Anderson,Guralnik,EnglertBrout,Higgs} as well as 
adjoint, theoretically emergent \cite{HofmannBook2016} gauge-symmetry breaking. 

Yang-Mills thermodynamics, a field theory formulated 
solely in terms of gauge potentials which live on the Euclidean cylinder $S_1\times {\bf R}^3$ of temporal extent 
$0\le x_4\le\beta=\frac{1}{T}$, even when subjected to the simplest non-Abelian group SU(2) comprises
complex ground states in each of its three phases -- deconfining, preconfining, and confining \cite{Hofmann2005}. 
In particular, the thermal ground state of the deconfining phase is represented by 
(anti)selfdual gauge-field configurations of topological charge modulus unity 
and trivial holonomy -- Harrington-Shepard (anti)calorons \cite{HS1977} -- 
whose spatially densely packed centers and overlapping peripheries determine the nature of this ground state's 
quantum and classical excitations, respectively \cite{GrandouHofmann,HofmannBook2016}. 
While (anti)caloron peripheries provide (anti)selfdual dipole fields, effectively creating electric and magnetic 
dipole densities whose undulating re-polarisations associate with the classical propagation of 
low-frequency electromagnetic waves \cite{GrandouHofmann}, (anti)caloron centers associate with thermal 
quasiparticle fluctuations which are short-lived and local quantum, that is, indeterministic 
excitations whose energy and momentum are governed by the (anti)caloron action (Planck's quantum of action $\hbar$ \cite{KavianiHofmann2012,HofmannBook2016}). Depending on whether quantum excitations occur {\sl within} or {\sl off} the Cartan subalgebra, defined by the direction of the effective, adjoint, inert, and spatially homogeneous 
scalar field $\phi$ in the SU(2) algebra su(2), they 
are {\sl massless} or exhibit a substantial {\sl mass  gap} (adjoint Higgs mechanism), respectively \cite{HofmannBook2016}. 
Notice that the field $\phi$ represents (anti)caloron centers (without packing voids) in a point-like way 
at a spatial resolution set by its own modulus \cite{HofmannBook2016}. The latter, in turn, is determined by an 
integration constant (the Yang-Mills mass scale $\Lambda$) and temperature $T$. Apart from small 
correlations, caused by {\sl boundary overlaps}/ {\sl packing voids} of centers 
as well as peripheries-enabled low-frequency waves and effectively mediated through well-controlled radiative 
corrections \cite{HofmannBook2016,BischerGrandouHofmann}, the quantum physics 
{\sl within} a given (anti)caloron center is independent of the quantum physics 
{\sl within} any other (anti)caloron center. This is the reason why the associated 
fluctuations, whose spatial range does not exceed $|\phi|^{-1}$, are subject to a 
Bose-Einstein distribution function. In contrast to the massive sector, whose excitations 
thus can only fluctuate within each (anti)caloron center separately, the massless 
sector propagates in a wave-like way for frequencies deeply within the Rayleigh-Jeans 
regime (the separation between photon-like, localized fluctuations and 
wave-like long-range propagation is set by a frequency scaling like $T^{-2}$ 
\cite{GrandouHofmann}). Therefore, information on temperature 
perturbations, carried by the massless sector, are effectively 
propagated at the speed of light (the average behaviour of localized photonic fluctuations 
riding the low-frequency waves of the deeply Rayleigh-Jeans regime). 
Massive fluctuations, on the other hand, do not contribute to the propagation of 
temperature fluctuations.                    

The present contribution sketches essential steps in the 
derivation of a useful a priori estimate of the deconfining thermal 
ground state, its emergent structure -- determinining its excitability in terms of waves 
and thermal quasiparticles --, and a number of applications to the Cosmic 
Microwave Background (CMB). In Sec.\,\ref{Sec1} we 
review properties of topological-charge-modulus-unity (anti)calorons of 
trivial (HS) and nontrivial (LLvBK) holonomy along with a discussion of their stability. 
How the time dependence of the field strength of HS (anti)caloron centers can 
be trivialised (to become a mere, time dependent choice of gauge) by a spatial coarse-graining over its adjointly 
transforming two-point correlator is discussed in Sec.\,\ref{Sec2}. 
In Sec.\,\ref{Sec3} we review the effective action 
after coarse-graining, characterized by an inert and adjoint field $\phi$,  
which breaks SU(2) to U(1), and also coarse-grained gauge fields,  
collectively describing (anti)caloron peripheries, their wave-like excitation, 
and localized thermal quasiparticle fluctuations which associate 
with (anti)caloron centers. This theory admits a loop 
expansion in a completely fixed and physical gauge with no need to introduce 
ghost fields to reduce the redundancy of perturbative field configurations under 
linear gauge conditions. Sec.\,\ref{Sec4} addresses the radiative 
corrections to the pressure and the polarisation tensor of 
the massless mode. Concerning the former, the counting of 
excluded scattering channels and energy-sign combinations of loop momenta vs. a priori 
possible ones in low-loop-order 2-particle-irreducible bubble diagrams for the pressure, 
arising solely from massive quasiparticle fluctuations, is performed. Next, an analytical computation of 
the high-$T$ behaviour of the three-loop diagram is 
sketched and confronted with numerical results for low $T$. At high $T$ this diagram 
grows with a power $T^{13}$, and therefore, by itself, does not represent a small correction 
to the Stefan-Boltzmann limit exhibited by free quasiparticle excitations (one-loop result). 
This motivates the resummation of all bubble diagrams with dihedral symmetry, the three-loop diagram being the 
lowest loop order. Together with numerical results 
for the two-loop diagram this resummation is carried out in terms of the 
associated Dyson-Schwinger equation, and a well controlled extrapolation to high-$T$ of the 
low-$T$ behaviour, the latter indicated by the three-loop diagram, is thus obtained. We also review earlier results on the one-loop polarisation tensor of the massless mode and its resummation to yield dispersion laws for 
longitudinally and transversely propagating disturbances. In Sec.\,\ref{Sec5} 
we discuss implications for the Cosmic Microwave Background (CMB) once the assumption is made 
that this thermal photon gas is subject to deconfining SU(2) 
rather than U(1) gauge dynamics. In particular, we emphasize how the Yang-Mills scale of this theory is fixed by radio-frequency CMB observations, the occurrence of a modified temperature-redshift 
relation in an FLRW Universe, its implications for the cosmological model at high redshifts with a potential resolution of 
the present discrepancy in $H_0$ when extracted from local cosmology and the CMB, 
radiatively induced anomalies for two-point correlation function at large angles, and the late-time emergence of intergalactic magnetic fields. Finally, in Sec.\,\ref{Sec6}, we summarise our present discussion and provide an outlook on future 
work. Most definitely, this concerns the resummation of dihedral diagrams involving massless and 
massive modes, the match between the successful low-redshift $\Lambda$CDM cosmological model and the new model 
at high redshifts in terms of percolated and depercolated vortices of a Planck-scale axion, and the according check 
of viability against the observed angular power spectra of CMB temperature and polarisation correlation functions with a possible dynamical clarification of large-angle anomalies.

\section{(Anti)calorons}
\label{Sec1}

The fundamental (Euclidean) SU(2) Yang-Mills action at finite temperature $T$ is 
\eqb
\label{fYact}
S=\frac{1}{2g^2}\mbox{tr}\int_0^\beta dx_4 d^3x\,F_{\mu\nu}F_{\mu\nu}\,,
\eqe
where the field-strength tensor $F_{\mu\nu}$ is defined as 
$F_{\mu\nu}=\pd_\mu A_\nu-\pd_\nu A_\mu+i[A_\mu,A_\nu]$ and the gauge field $A_\mu$ as $A_\mu=A_\mu^a t_a$ ($\mu, \nu=1,2,3,4$, $a=1,2,3$, 
and $A_\mu^a$ real), tr demands tracing over products of matrices in su(2), 
$t^a$ are hermitian generators (a basis in su(2)) in the fundamental 
representation ($2\times 2$ matrices), which are normalised as tr\,$t_a t_b=\frac12 \delta_{ab}$, 
and $g$ is the fundamental gauge coupling. (Anti)selfdual 
configurations ($F_{\mu\nu}=\pm\frac12 \epsilon_{\mu\nu\kappa\lambda} F_{\kappa\lambda}$ with 
$\epsilon_{1234}=1$ and totally antisymmetric) solve the second-order Yang-Mills equations 
$D_\mu F_{\mu\nu}=0$ (covariant, adjoint derivative defined as $D_\mu \cdot=\pd_\mu\cdot +i[A_\mu,\cdot]$) 
and, in a given topological sector (homotopy group $\Pi_3(\mbox{SU(2)})={\bf Z}$) of topological charge $k\in {\bf Z}$ 
saturate the Bogomoln'yi bound for the Yang-Mills action of Eq.\,(\ref{fYact}): 
\eqb
\label{Bogo}
S=\frac{8\pi^2}{g^2}|k|\,.
\eqe
As one can easily show \cite{ShuryakSchaefer,HofmannBook2016}, (anti)selfdual 
gauge-field configurations cause the energy-momentum tensor $\theta_{\mu\nu}$ of 
the theory to vanish identically which makes them candidates for 
the composition of the thermal ground state. 

The venue for the construction of periodic configurations with $|k|=1$ was opened by 't Hooft 
\cite{tHooft1975} who showed that in the so-called singular gauge, 
where the $|k|=1$ instanton configuration $A_\mu$
on ${\bf R}^4$ is singular at the peak of its action density \cite{JackiwRebbi1975,tHooft1975}, the gauge 
field $A_\mu$ for $|k|>1$ can be represented as
\eqb
\label{singgauge}
A_\mu=\left\{\begin{array}{c}\bar{\eta}^a_{\mu\nu} t_a\pd_\nu\log\Pi(x)\ \ \ \ (k>0)\\ 
\ \ \eta^a_{\mu\nu} t_a\pd_\nu\log\Pi(x)\ \ \ \ (k<0)\,.\end{array}\right.
\eqe
Here, the scalar function $\Pi$ (pre-potential for $k>1$) is a superposition 
of gauge-field pre-potentials for $k=1$. In Eq.\,(\ref{singgauge}) $\bar{\eta}^a_{\mu\nu}$ and $\eta^a_{\mu\nu}$ 
denote the antisymmetric-in-$\mu\nu$ and (anti)selfdual 't Hooft symbols (required when decomposing a pure-gauge 
configuration, winding with $|k|=1$ at spacetime infinity, into its su(2) components). This superposition 
principle for the pre-potential on ${\bf R}^4$ was exploited by 
Harrington and Shepard \cite{HS1977} to construct an infinite "mirror sum", based on a "seed" pre-potential  
centered at $\mathbf{x}=x_4=0$ and of instanton scale parameter $\rho$ (a measure for  
the spread of action density), to achieve periodicity in $x_4$ and 
topological charge $|k|=1$ on $S_1\times {\bf R}^3$. They obtain 
\eqb
\label{eomWinsolbetaexp}
\Pi(x_4,r;\rho,\beta)=
1+\frac{\pi\rho^2}{\beta r}
\frac{\sinh\left(\frac{2\pi r}{\beta}\right)}{\cosh\left(\frac{2\pi r}{\beta}\right)-
\cos\left(\frac{2\pi x_4}{\beta}\right)}\,,
\eqe
where $r\equiv |\mathbf{x}|$. It is easy to check 
that for $r\to\infty$ the exponential of the line integral of 
$A_4$ along $x_4$ (the Polyakov loop) is unity: this configuration 
is of trivial holonomy. We also note that the function $\Pi$ 
in Eq.\,(\ref{eomWinsolbetaexp}) approximates as follows for 
$|x|\equiv \sqrt{x_\mu x_\mu}\ll\beta$ and for $r\gg\beta$ \cite{GPY}:
\eqb
\label{limitsPi}
\Pi=\left\{\begin{array}{c}\left(1+\frac13\frac{s}{\beta}\right)+\frac{\rho^2}{x^2}\ \ \ \ (|x|\ll\beta)\\ 
\ \ \ 1+\frac{s}{r}\ \ \ \ \ \ \ \ \ \ \ \ \ \ \ (r\gg\beta)\,,\end{array}\right.
\eqe
where the length scale $s$ is given as $s=\frac{\pi\rho^2}{\beta}$. 
For $|x|\ll\beta$ the field strength $F_{\mu\nu}$ is that of an (anti)instanton with a re-scaled parameter ${\rho^\prime}^2=\frac{\rho^2}{1+\frac{s}{\beta}}$, for $\beta\ll r\ll s$ one has
\eqb
\label{monoBE}
B^a_i\equiv\frac12\epsilon_{ijk}F^a_{jk}=\pm E^a_i\equiv\pm F^a_{4i}=
\frac{\hat{x}^a\hat{x}_i}{r^2}\,,
\eqe
and for $\beta\ll s\ll r$
\eqb
\label{monoDi}
B^a_i=\pm E^a_i=s\frac{\delta_i^a-3\hat{x}_i\hat{x}^a}{r^3}\,,
\eqe 
where $\hat{x}_i\equiv \frac{x_i}{r}$ and $\hat{x}^a\equiv \frac{x^a}{r}$. 
Since, as we will argue in Sec.\,\ref{Sec2}, the value of the scale parameter 
$\rho$ is sharply centered at $|\phi|^{-1}$ for all those HS (anti)calorons 
that contribute to the a priori estimate of the thermal ground state of the deconfining 
phase and with a lower bound $T_c$ on possible temperatures, proportional to the Yang-Mills scale $\Lambda$, 
one easily shows \cite{GrandouHofmann} that the following hierarchy is always satisfied:
\eqb
\label{hier}
\beta\ll |\phi|^{-1}\ll s\,,
\eqe  
which makes the above considered spatial distance regimes, implying Eqs.\,(\ref{monoBE}) ((anti)selfdual, static monopole) 
and (\ref{monoDi}) ((anti)selfdual, static dipole), relevant. 

At $|k|=1$ (anti)calorons with nontrivial holonomy (Polyakov loop at spatial infinity not equal to an element of the SU(2) center group $\{-{\bf 1},{\bf 1}\}$) are much harder to construct \cite{LeeLu,KraanVanBaal} via Nahm's beautiful and deep 
transformation between (anti)selfdual gauge-field configurations on a torus and its dual torus \cite{Nahm}. Nontrivial 
holonomy can be prescribed, e.g., by letting $A_4(r\to\infty,x_4)=ut_3$ with 
$0<u<\frac{2\pi}{\beta}$. Thus $A_4$ must be considered an adjoint Higgs field for the spatial components $A_i$, 
inducing, at overall charge 
neutrality, a pair of a static magnetic monopole and its antimonopole with an exact cancellation between the $A_4$-field mediated repulsion and the $A_i$-field mediated attraction. This situation can smoothly be 
connected to trivial holonomy $u\to 0$ or $u\to\frac{2\pi}{\beta}$. Namely, given $\rho$ 
and $\beta$ the monopole-antimonopole pair is spatially separated by the length scale $s$, and the holonomy 
$u$ assigns masses to them as $m_m=4\pi u$ and $m_a=4\pi\left(\frac{2\pi}{\beta}-u\right)$ ($g=1$). Therefore, trivial 
holonomy renders one of these defects massless, spreading it over space, while its 
partner remains massive and localised. (Notice that masslessness for both monopole and antimonopole 
can also be enforced at nontrivial holonomy if the gauge coupling is sent to infinity 
($g\to\infty$) in which case both defects become point-like. This can be seen by absorbing $g^{-1}$ into the gauge fields of Eq.\,(\ref{fYact}), see also \cite{tHooft6,PrasadSommerfield}.)  

As we will argue in Sec.\,\ref{Sec2}, only (anti)calorons with $|k|=1$ may enter the a priori estimate of the thermal ground state. Accepting this for the moment, the question remains how the holonomy is to be picked. 
By a heroic deed, Diakonov and collaborators showed in \cite{Diakonov} by computing the 
one-loop quantum weights along the lines of 't Hooft's calculation of the instanton weight \cite{tHooft1975} that nontrivial 
holonomy makes the configuration $A_\mu$ unstable. Namely, for $\frac{s}{\beta}=\pi\left(\frac{\rho}{\beta}\right)^2\gg 1$, which is a relevant limit, see (\ref{hier}), small holonomy ($0\le u\le \frac{\pi}{\beta}(1-\frac{1}{\sqrt{3}})$ and 
$\frac{\pi}{\beta}(1+\frac{1}{\sqrt{3}})\le u\le 2\,\frac{\pi}{\beta}$) perturbes the 
balance between attraction and repulsion in favour of attraction, while the opposite is true 
of the situation with large holonomy (complementary range), causing the (anti)caloron to dissociate. 
As it is easily argued \cite{HofmannBook2016}, the former situation is much more likely  
than the latter one. To construct an a priori estimate, however, neither large nor 
small holonomy are admissible due to the instability of the associated (anti)caloron. This only 
leaves the HS (anti)caloron as a valid component of the thermal ground-state estimate.     
 
\section{Spatial coarse-graining and a two-point correlator}
\label{Sec2}

Let us now quote the essential steps in deriving the field $\phi$ from a spatial 
coarse-graining over the following two-point function: 
\eqb
\{\hat{\phi}^a\}\equiv\sum_{C,A}{\rm tr}\!\!\int d^3x \int d\rho \,t^a\,
F_{\mu\nu} (\tau,\mathbf{0}) \,\{(\tau,\mathbf{0}),(\tau,\mathbf{x})\} F_{\mu\nu} (\tau,\mathbf{x}) \{(\tau,\mathbf{x}),
(\tau,\mathbf{0})\}\,,
\label{definition}
\eqe
where 
\begin{equation}
\label{abk}
\{(\tau,\mathbf{0}),(\tau,\mathbf{x})\}\equiv
{\cal P} \exp \left[ i \int_{(\tau,\mathbf{0})}^{(\tau,\mathbf{x})} dz_{\mu} \, A_{\mu}(z) \right]\,,\ \ 
\{(\tau,\mathbf{x}),(\tau,\mathbf{0})\}\equiv\{(\tau,\mathbf{0}),(\tau,\mathbf{x})\}^\dagger\,,
\end{equation}
and, from now on, $\tau\equiv x_4$. The Wilson lines in Eq.\,(\ref{abk}) are calculated along
the straight spatial line connecting the points $(\tau,\mathbf{0})$ and $(\tau,\mathbf{x})$, and
${\cal P}$ demands path-ordering. In (\ref{definition}) the sum is over the $|k|=1$ HS 
caloron ($C$) and anticaloron ($A$), and $\{\hat{\phi}^a\}$ signals a family of 
(dimensionless) phases of the field $\phi$ whose continuous parameters 
emerge partially in the course of evaluating the the right-hand side and partially relate to temporal shift moduli. It is straight-forward to 
argue \cite{HofmannBook2016} that (\ref{definition}) is unique: adjointly transforming 
one-point functions vanish identically due to (anti)selfduality, higher $n$-point functions and higher topological charges are excluded by dimensional counting, the coincidence of the spatial (anti)caloron center 
with $\mathbf{0}$ is demanded by spatial isotropy, and the straight-line evaluation of Wilson 
lines by the absence of any spatial scale on the classical (Euclidean) level. Actually, 
this allows to ignore the path-ordering prescription since $A_i$ is a spatial hedge-hog, centered at $\mathbf{0}$, 
which assigns to each {\sl direction} in ${\bf R}^3$ the same direction in su(2). 

As a result of performing the integrations in (\ref{definition}) it turns out that 
$\{\hat{\phi}^a\}$ is characterized by harmonic motion of period $\beta$ 
within some plane ${\bf R}^2\subset \mbox{su(2)}$ (global gauge choice)
subject to unspecified normalizations and phases for the oscillations along each of 
the two axes \cite{HerbstHofmann2004}. Therefore, $\{\hat{\phi}^a\}$ uniquely 
comprises the kernel of the {\sl linear} second-order operator ${\cal
D}\equiv\partial^2_\tau+\left(\frac{2\pi}{\beta}\right)^2$. Because (anti)caloron action 
is independent of temperature, however, the explicit temperature dependence in ${\cal
D}$ must be absorbed into the $\phi$-derivative of a potential $V$. Demanding $\phi$ to be simultaneously 
the solution of a first-order BPS and the Euler-Lagrange equation, one derives the following first-order equation 
for $V$ \cite{HofmannBook2016}
\begin{equation}
\label{eomPot}
\frac{\partial V(|\phi|^2)}{\partial
  |\phi|^2}=-\frac{V(|\phi|^2)}{|\phi|^2}
\end{equation}
with solution 
\begin{equation}
\label{solPot}
V(|\phi|^2)=\frac{\Lambda^6}{|\phi|^2}\,,
\end{equation}
where $\Lambda$ denotes an arbitrary mass scale (the Yang--Mills
scale). Since the BPS equation, which exhibits the {\sl square root} of $V$ and demands "circular polarisation" for the harmonic motion in the plane, needs to be satisfied 
in addition to the Euler-Lagrange equation (first derivative of $V$) the usual additive 
shift symmetry of the potential in the Euler-Lagrange equation no 
longer is an option: Should $\pm V=\pm 4\pi\Lambda T^3$ (with $|\phi|=\sqrt{\frac{\Lambda^3}{2\pi T}}$) 
alone turn out to represent a good a priori estimate of the thermal ground state's energy density and pressure, respectively (see Sec.\,\ref{Sec3}), 
then this result is unique.  

It is important to note that the integration over the instanton scale parameter 
$\rho$ depends {\sl cubically} on an upper integration limit $\rho_u$ 
and that the dependence on $\tau$ of this integral saturates very 
rapidly for $\rho_u/\beta>1$ into the harmonic one. Therefore, the kernel $\{\hat{\phi}^a\}$ 
in Eq.\,(\ref{definition}) of the differential operator ${\cal D}$ is strongly dominated by a small band of $\rho$ 
values centered at the cutoff $\rho_u=|\phi|^{-1}$, and one can show 
that $\rho_u/\beta=|\phi|^{-1}/\beta\gg 1$ for all temperatures within the 
deconfining phase \cite{HofmannBook2016}.

\section{Effective action, adjoint gauge-symmetry breaking, waves, and 
thermal quasiparticles}
\label{Sec3}

Since $\phi$ is inert (no momentum transfer to and from this field) 
and the Yang-Mills action, restricted to topologically trivial gauge fields, 
is renormalisable \cite{tHooftVeltman} and since the effective 
Lagrangian density ${\cal L}_{\rm eff}$ (after spatial coarse-graining) is required 
to be gauge invariant, one arrives at the following, unique answer \cite{HofmannBook2016}
\eqb
\label{actdeneff}
{\cal L}_{\rm eff}[a_\mu]={\rm tr}\,\left(\frac12\,
  G_{\mu\nu}G_{\mu\nu}+(D_\mu\phi)^2+\frac{\Lambda^6}{\phi^2}\right)\,,
\eqe
where $G_{\mu\nu}=\partial_\mu a_\nu-\partial_\nu
a_\mu-ie[a_\mu,a_\nu]\equiv G^a_{\mu\nu}\,t_a$ denotes the field
strength of the {\it effective} trivial-topology gauge field $a_\mu=a_\mu^a\,t_a$,
$D_\mu\phi=\partial_\mu\phi-ie[a_\mu,\phi]$, and $e$ is the effective
gauge coupling. A solution $a_\mu^{\rm gs}$ to the effective, second order Yang-Mills 
equations with $D_\mu\phi=G_{\mu\nu}=0$ reads $a_\mu^{\rm gs}=
\delta_{\mu 4}\frac{2\pi}{e\beta}\,t_3$ if without restriction of generality (global gauge choice) 
circular polarisation in the 1-2 plane of su(2) is considered for $\phi$. On $\phi$ and 
$a_\mu^{\rm gs}$ the action density (\ref{actdeneff}) thus reduces to the potential $V$, and 
the vanishing energy density and pressure of (anti)caloron centers is made finite by (anti)caloron overlap, 
effectively represented by $a_\mu^{\rm gs}$:
\eqb
\label{gsrhoPress}
\rho^{\rm gs}=-P^{\rm gs}=4\pi\Lambda^3 T\,.
\eqe   
On $a_\mu^{\rm gs}$ the Polyakov loop turns out to be $-{\bf 1}$, and one can show that a 
singular but admissible (time periodic) gauge rotation exists which transforms $\phi$ to unitary 
gauge $\phi=2\,|\phi|\,t_3$ and $a_\mu^{\rm gs}$ to $a_\mu^{\rm gs}=0$ where the Polyakov loop now is 
${\bf 1}$. This demonstrates the electric ${\bf Z}_2$ degeneracy of the thermal ground-state estimate -- a sure sign 
of deconfinement. 

In unitary gauge one can read off from the action density (\ref{actdeneff}) the mass spectrum 
$m^2_a=-2e^2{\rm tr}\,[\phi,t_a][\phi,t_a]$ for the gauge-field 
excitations as induced by the adjoint Higgs mechanism:  
\begin{equation}
  \label{massessu2}
  \begin{split}
    &m^2 \equiv  m_1^2=m_2^2=4e^2\frac{\Lambda^3}{2\pi T},\\[3pt]
    &m_3 = 0\,.
  \end{split}
\end{equation}
If, instead, one appeals to the Dyson series for mass generation then one can easily 
show that any attempt to move a massive gauge-field fluctuation away from its mass 
shell would inevitably transfer energy-momentum to the field $\phi$. This, however, would 
contradict the very derivation of $\phi$. Thus massive quasiparticle fluctuations, albeit deeply 
probing and thus orginated by (anti)caloron centers (quantum excitation, no classical or quasi-classical 
wave propagation: $e\ge\sqrt{8}\pi$ $\Rightarrow$ $m=\omega\gg |\phi|$, see below, where 
$\omega$ is the circular would-be frequency of a wave-like propagation, compare with \cite{GrandouHofmann} for the massless case $m_3=0$) need to be understood 
as Bose-Einstein distributed quasiparticle on-shell (quantum) excitations.   

By imposing thermodynamical selfconsistency onto the thermal ground-state estimate and its free thermal 
quasiparticle excitations (intactness of Legendre transformation between thermodynamical 
quantities as computed in the theory given by Eq.\,(\ref{actdeneff})) one derives the following evolution equation for the effective gauge coupling $e$   
\begin{equation}
\label{evalambdasu2}
1=-\frac{24\lambda^3}{(2\pi)^6}\left(\lambda\frac{da}{d\lambda}+a\right)a\,D(2a)\,,
\end{equation}
where $D(y)\equiv\int_0^\infty dx\,\frac{x^2}{\sqrt{x^2+y^2}}\frac{1}{\e^{\sqrt{x^2+y^2}}-1}$, $a\equiv\frac{m}{2T}$, and $\lambda \equiv \frac{2\pi T}{\Lambda}$. Lowering $\lambda$ from a sufficiently high initial value $\lambda_i$ the solutions to Eq.\,(\ref{evalambdasu2}), when solved for $e(\lambda)$, linearly fast run into the attractor 
$e=\sqrt{8}\pi$ for $\lambda_i\gg\lambda\gg\lambda_c=13.87$ and $e\propto -\log(\lambda-\lambda_c)$ 
for $\lambda\sim\lambda_c$ \cite{HofmannBook2016}. Since only (anti)calorons with $\rho\sim |\phi|^{-1}$ 
contribute to the emergence of field $\phi$ one can use $g=e$ in the (anti)caloron action (\ref{Bogo}) 
($|k|=1$). Moreover, by re-instating $\hbar$ in the effective action, one easily arrives at $e=\frac{\sqrt{8}\pi}{\sqrt{\hbar}}$ \cite{Brodsky,KavianiHofmann2012,HofmannBook2016}, and thus $S=\hbar$. 
At $\lambda_c$ the magnetic (anti)monopoles liberated by rarely occurring dissociations of large-holonomy (anti)calorons -- an effectively described radiative effect \cite{LudescherHofmann,SpatialWilson,HofmannBook2016} -- become point-like and massless.

For the estimate of thermodynamical quantities based on free thermal quasiparticle fluctuations to be reliable, it must be assured that radiative (nonthermal) corrections remain small. The next Section provides results indicating that, modulo possible 
resummations, this indeed is the case. 

\section{Radiative corrections: Massive sector}
\label{Sec4}

Here we only consider higher loop corrections involving massive thermal quasiparticle 
fluctuations. The discussion is a summary of the results obtained 
in \cite{BischerGrandouHofmann} and shows that, in general, radiative 
corrections to the pressure in the deconfining phase, although small and well under control for all temperatures, 
are not subject to a thermodynamical interpretation.     

\subsection{Counting excluded vs. allowed scattering-channel and energy-sign 
combinations in bubble diagrams}

In \cite{BischerGrandouHofmann} a systematic study of 2-particle-irreducible (2PI) bubble 
diagrams with massive lines only, which radiatively contribute to the loop expansion of the pressure in the deconfining phase, 
was performed. The region of integration over loop momenta is severely 
constrained by the demand that in each scattering channel for a given 
four-vertex the invariant momentum transfer (Mandelstam variables $s$, $t$, and $u$) is bounded 
by $|\phi|^2$ in unitary-Coulomb gauge. A measure of how constrained loop momenta turn out to be is obtained by 
comparing the a priori (without constraints) possible combinations of their energy signs in a given diagram 
with the allowed combinations after imposing the constraints.  The ratio $R(L)$ 
of the latter to the former in dependence of loop order $L$ maximally is 
\eqb
\label{LoopOrderR}
R(3)=0.1667\,,\ \ R(4)=0.0463\,,\ \ R(5)=0.0123\,,\ \ R(6)=0.0044\,.
\eqe
Multiplying $R$ with the symmetry factor $S$ of a diagram maximally yields
\eqb
\label{LoopOrderRS}
R\cdot S(3)=0.00347\,,\ \ R\cdot S(4)=0.000965\,,\ \ R\cdot S(5)=0.000386\,,\ \ R\cdot S(6)=0.000113\,.
\eqe
Eqs.\,(\ref{LoopOrderR}) and (\ref{LoopOrderRS}) testify of an impressive suppression 
of allowed loop fluctuations with increasing loop order as imposed by the vertex constraints. 
Yet, one can demonstrate that diagrams with dihedral symmetry (polygons inscribed into the circle with corners defining 
four-vertices) will exhibit allowed energy-sign combinations at an arbitrarily high 
loop order. This motivates the explicit computation of the dihedral diagram with $L=3$. 

\label{Sec41}
\subsection{Low- and high-temperature behavior of three-loop diagram}
\label{Sec42}

\begin{figure}[h]
\centering
\includegraphics[width=3cm,clip]{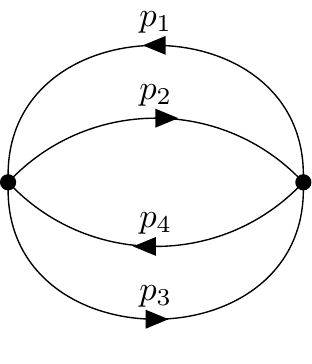}
\centering
\caption{Dihedral diagram with $L=3$.}
\label{fig-1} 
\end{figure}

The three-loop diagram in Fig.\,\ref{fig-1} is subject to the following 
expression \cite{BischerGrandouHofmann}
\begin{multline}\label{eq:3-loop-start}
\Delta P|_{\mathrm{3-loop}} = i\frac{\Lambda^4}{48\lambda^2}e^4\frac1{(2\pi)^6}\sum_{\rm{signs}}
\int d\theta_1d\varphi_1dr_1dr_2d\theta_3\sum_{\{r_3\}}r^2_1r^2_2r^2_3\sin\theta_1\sin\theta_3\\
\times P(p_i)\frac{n_B'(r_1)n_B'(r_2)n_B'(r_3)n_B'(r_4)}{8|p^0_1p^0_2p^0_3p^0_4|}\,.
\end{multline}
The first sum in \eqref{eq:3-loop-start} runs over allowed sign combinations for 
$p^0_i$, $i=1,\dots,4$, we have scaled all momentum components $p_{i}^\mu$ and mass $m$ dimensionless by $|\phi|$, 
$n_B'(r_i)\equiv n_B\left(2\pi\sqrt{\frac{r_i^2+m^2}{\lambda^3}}\right)$ ($n_B$ the Bose-Einstein distribution),   
$P(p_i)$ denotes a certain polynomial of invariants formed by the loop four-momenta $p_i$, $r_i$ denotes the modulus of 
the spatial momentum $\mathbf{p}_i$, and angular variables parametrise their mutual orientation.    
The second sum in \eqref{eq:3-loop-start} runs over all solutions in $r_3$ of the equation
\begin{multline}\label{eq:r3}
\sgn(p^0_2)\sqrt{r_2^2+m^2}+\sgn(p^0_3)\sqrt{r^2_3+m^2}-\sqrt{r_1^2+m^2}=
-\left[r^2_1+r^2_2+r^2_3 \right.\\\left. - 2 r_1r_2\cos\theta_1  -2r_1r_3(\sin\varphi_1\sin\theta_1\sin\theta_3+\cos\theta_1\cos\theta_3)+2r_2r_3\cos\theta_3 + m^2\right]^{1/2}\,.
\end{multline}
While an analytical treatment of \eqref{eq:3-loop-start} is impossible at 
low temperatures and therefore requires Monte-Carlo integration one can, 
however, extract the leading power in $\lambda$ in a 
high-temperature expansion. Namely, one obtains 
\eqb
\begin{split}\label{eq:approxintss3}
\Delta P|_{\mathrm{3-loop}}&= i\frac{\Lambda^4}{\lambda^2}e^4\frac1{(2\pi)^5}\frac1{15}\left(1+\frac{1}{4m^2}\right)\frac{I_3^2}{m^8}\\
&=i\Lambda^4\frac{1}{3375}\frac{1}{(2\pi)^{15}}\frac{1}{m^4}\left(1+\frac{1}{4m^2}\right)\left(\pi^4-90\zeta(5)\right)^2\lambda^{13}\\ 
&=5.2968\cdot10^{-20}i\Lambda^4\lambda^{13}\,.
\end{split}
\eqe
Eq.\,(\ref{eq:approxintss3}) is the result of an elaborate analysis which is confirmed 
by the numerical computation. The latter predicts an impressive downward hierarchy between one-loop, two-loop, and three-loop 
contributions at small temperatures. 
The power of thirteen at high temperatures is strongly indicative that a resummation of all dihedral loop 
orders is required since the Stefan-Boltzmann power of the one-loop contribution to the pressure exhibits a power 
of four only. Notice that $\Delta P|_{\mathrm{3-loop}}$ is dominated by a purely imaginary contribution at high 
temperatures.   

\subsection{Resummation of dihedral diagrams}
\label{Sec43}

The resummation of dihedral diagrams to all loop orders is performed in terms of the solution of the according Dyson-Schwinger equation for the vertex form factor $f(\lambda)$ at high temperatures. One obtains \cite{BischerGrandouHofmann}
\eqb
\label{eq:3lresummed}
\begin{split}
f^2(\lambda)\Delta P|_\mathrm{3-loop}&\approx \left(-0.94\cdot 10^{15}i\lambda^{-11.6}\right)^2\cdot5.3\cdot 10^{-20}i\Lambda^4\lambda^{13}\\
&=-4.7\cdot 10^{10}i\Lambda^4\lambda^{-10.2}\,.
\end{split}
\eqe 
Thus, upon resummation the power of thirteen of the "naked" three-loop diagram 
has dropped to a power of roughly minus ten as a result of resummation. One can also show 
that the leading powers of the resummed two-loop and three-loop diagrams cancel. 
In general, this does not imply the cancellation of subleading powers, and one 
should count on the appearance of imaginary contributions. Physically, these introduce 
small nonthermal behavior (turbulences) which can be envisaged corrections 
to the isotropic and homogeneous a priori estimate of the thermal ground-state field $\phi$ in terms 
of a single (anti)caloron center as introduced by slight overlaps and packing voids.       

\section{Radiative corrections: Massless sector}

In addition to an analysis of feeble one-loop photon-photon scattering in \cite{Krasowski} and 
the two-loop bubble diagrams involving massless modes \cite{Schwarz}, which contribute to the pressure,  
we have computed the polarisation tensor of the massless mode at resummed one-loop 
order in \cite{LudescherHofmann,FalquezHofmannBaumbachI} and in \cite{FalquezHofmannBaumbachII}. This requires the solution of 
gap equations for the screening functions $G$ and $F$ for transverse and longitudinal photons, 
respectively. Briefly, function $G$ predicts a low-frequency gap (screening) in the spectral intensity (radiance) of a 
black body which closes like $\lambda^{-1/2}$ with increasing temperature and is absent at 
$\lambda_c$. Shortly above the gap a localised spectral region of antiscreening is encountered. 
Function $F$, on the other hand, associates with three branches of low-momentum longitudinal 
modes (charge-density waves).     

\section{Implications for the Cosmic Microwave Background (CMB) and the Cosmological Model}

\label{Sec5}

There are various implications of deconfining SU(2) Yang-Mills thermodynamics for the CMB once the postulate 
is made that it describes thermal photon gases with an electric-magnetic dual 
interpretation of its Cartan subalgebra \cite{HofmannBook2016}. The latter property is easily derived from $e=\frac{\sqrt{8}\pi}{\sqrt{\hbar}}$, see Sec.\,\ref{Sec3}, and the fact that in the same system of units ($c=1=\epsilon_0=\mu_0$) the fine-structure 
constant of QED reads $\alpha=\frac{Q^2}{4\pi\hbar}$ since this implies that the electric charge $Q$ is inversely proportional 
to the effective Yang-Mills coupling $e$. Since $g=\frac{4\pi}{e}$ is the charge of a magnetic 
monopole in SU(2) Yang-Mills thermodynamics this means that electric charges in the real world are magnetic charges w.r.t U(1)$\subset$SU(2) and vice versa.     

In particular, such a description should apply 
to the CMB. In \cite{Hofmann2009} an excess of spectral CMB power 
detected from about 3\,GHz \cite{Arcade2} down to 40\,MHz \cite{terrestialobserv} was interpreted as 
the onset of electric monopole condensation and the associated Meissner 
effect (evanescence of low-frequency waves). This, however, can only be the case if $T_c$ 
practically coincides with the present CMB baseline 
temperature $T_0=2.725\,$K, setting the Yang-Mills scale 
to $\Lambda_{\rm CMB}\sim 10^{-4}\,$eV. 

Let us mention a few consequences. First, an SU(2) Yang-Mills theory of this scale 
implies a modified temperature ($T$)-- redshift ($z$) relation in an FLRW universe 
\cite{Hofmann2015}. Namely, there is curvature in this relation at low $z$, and at high $z$ one 
has $T/T_0=0.63(z+1)$. There are two immediate physics implications: (i) resolution of 
the re-ionisation puzzle (high $z_{\rm re}$ from extraction using the CMB angular 
power spectra, low $z_{\rm re}$ from detection of Gunn-Peterson trough in quasar spectra \cite{Becker}) \cite{Hofmann2015}, 
(ii) drastic change of cold matter content in high-$z$ cosmological model \cite{HahnHofmann} since 
recombination now occurs at a considerably higher redshift with a resolution of the $H_0$ (today's 
value of Hubble parameter) puzzle (low $H_0$ from extraction using the CMB angular 
power spectra, high $H_0$ from local cosmological observation (standard candles, 
comoving sound horizon at baryon drag in matter correlation functions, absolute distance 
calibrations \cite{Adam}; time structure of gravitationally lensed quasar light \cite{holicow}). Second, due 
to a nontrivial screening function $G$ a cosmologically local depression in the temperature 
distribution of the CMB is dynamically generated \cite{SzopaHofmann2007,LudescherHofmannDep,HofmannNaturePhysics2013} 
which affects the CMB dipole (besides a kinematic contribution arising from the Doppler effect \cite{Peebles}), 
the low lying multipoles in the $TT$ and other correlation functions \cite{SchwarzHuterer} (large-angle anomalies), 
and could invoke the late-time emergence of intergalactic magnetic fields \cite{intgalmfields} through 
longitudinal modes, described by screening function $F$ \cite{FalquezHofmannBaumbachII}, subject to dynamical breaking of statistical isotropy \cite{HofmannNaturePhysics2013}.          

\section{Summary and Conclusions}
\label{Sec6}

This contribution to ICNFP 2017 has sketched the nonperturbative physics in the 
deconfining phase of SU(2) Yang-Mills thermodynamics and outlined a few applications 
towards the CMB, the cosmological model, and the late-time emergence of intergalactic magnetic fields. 
In particular, the computation of the CMB power 
spectra within the new cosmological model and taking into 
account radiative effects in SU(2) Yang-Mills thermodynamics (large-angle anomalies) is 
under way, and the "microscopic" physics (the relevant spatial distances are typical radii of spiral galaxies) 
for the percolation/depercolation transition of 
vortices in a Planck-scale axion field \cite{Giacosa,Neubert} should be understood. Such 
a transition is required to interpolate the new high-$z$ cosmological model to successful, 
low-$z$ $\Lambda$CDM cosmology \cite{HahnHofmann}.

\end{document}